# A Framework in CRM Customer Lifecycle: Identify Downward Trend and Potential Issues Detection

Kun Hu[1], Zhe Li[2], Ying Liu[1], Luyin Cheng[2], Qi Yang[2], and Yan Li[2]

## Abstract

Customer retention is one of the primary goals in the area of customer relationship management. A mass of work exists in which machine learning models or business rules are established to predict churn. However, targeting users at an early stage when they start to show a downward trend is a better strategy. In downward trend prediction, the reasons why customers show a downward trend is of great interest in the industry as it helps the business to understand the pain points that customers suffer and to take early action to prevent them from churning. A commonly used method is to collect feedback from customers by either aggressively reaching out to them or by passively hearing from them. However, it is believed that there are a large number of customers who have unpleasant experiences and never speak out. In the literature, there is limited research work that provides a comprehensive and scientific approach to identify these "silent suffers".

In this study, we propose a novel two-part framework: developing the downward prediction process and establishing the methodology to identify the reasons why customers are in the downward trend. In the first prediction part, we focus on predicting the downward trend, which is an earlier stage of the customer lifecycle compared to churn. In the second part, we propose an approach to figuring out the cause (of the downward trend) based on a causal inference method and semi-supervised learning. The proposed approach is capable of identifying potential silent sufferers. We take bad shopping experiences as inputs to develop the framework and validate it via a marketing A/B test in the real world. The test readout demonstrates the effectiveness of the framework by driving 88.5% incremental lift in purchase volume.

## Keywords

Customer Relationship Management, Semi-supervised learning, Customer Retention, Downward Prediction

[1] eBay Research & Engineering, Shanghai, China
[2] eBay e-Commerce, Shanghai, China
Email Address: kunhu@ebay.com (Kun Hu), zheli4@ebay.com (Zhe Li), yliu26@ebay.com (Ying Liu), lncheng@ebay.com (Luyin Cheng), qyang3@ebay.com (Qi Yang), yanli4@ebay.com (Yan Li)

# 1. Introduction

Customer retention is always one of the primary goals of the customer relationship management (CRM) field, because it brings direct value to the business (Chen and Popovich, 2003). However, in the customer lifecycle, some of the users churn for various reason. It is reasonable to try to "rescue" these churned customers since it is expensive to acquire new customers. An ideal approach to save customers is in the early stage, when they first shows a downward trend. To achieve this goal, two steps should be considered. The first one is to predict the propensity of the downward trend of a customer after a target length of time. Next, for a customer with a high propensity of a downward trend, we need to understand the cause of the issue. Depending on the cause, it is possible to execute personalized "rescue" plans – for instance, sending marketing messages, or providing incentives and relevant merchandise recommendations via multiple marketing channels.

There is some of the existing work using machine-learning methods which predict churned customers who stop doing business with a company for a certain period of time (Kamalraj and Malathi, 2013; Almana and Aksoy et al., 2014). Support vector machines are applied in the churn prediction of newspaper subscribers (Coussement and Van den Poel, 2008). Churn prediction is also widely used in telecommunication industry (Wei and Chiu, 2002; Lee and Lee et al., 2017), e-commerce retailers (Subramanya and Somani, 2017), social network (Óskarsdóttir and Bravo et al., 2017) and online new media (Yuan and Bai et al., 2017). In addition, methods are proposed to improve prediction accuracy; for instance, handling of imbalanced data, which usually happens in the customer relationship management field due to lack of positive cases (Burez and Van den Poel, 2009; Xie and Li et al., 2009). Although a large amount of research exists which concentrates on user churn, customer downward trend prediction is quite different as the customers are still active. Currently, no widely accepted definition exists that identifies the targeted customers with a downward trend. Lack of existing research design drives a necessity to develop a method for exploring the cause of downward trend.

In this work, we propose a framework to find the target population and causes, and making it possible for a company to address the issues implicitly or explicitly in their communication. We propose a scientific framework leveraging machine-learning methodology to manage the downward trend. To illustrate and demonstrate the methodology, we take an example of an e-commerce company.

The framework consists of three parts. First, a methodology is proposed to identify customers who are in the downward trend. Second, supervised learning models are built to identify the customers who are in a downward trend. Finally, we leverage semi-supervised machine learning to learn the potential causes, which later we call "levers" of the downward trend, and find silent sufferers. We will use one common lever - bad customer experience (BCE) to develop and demonstrate the method. Traditionally, if there are enough samples labeled as BCE sufferers and non-BCE sufferers, the issue is easy to handle with supervised learning method. However, not all of the customers are willing to express their BCE and thus they are silent sufferers. Meanwhile, BCE counts can be high if the user has more activities, and it is hard to establish the correlation between BCE and downward trend. Proposing a correct solution is crucial to rescuing these customers. In addition, we perform an A/B test to verify the performance of the model.

# 2. Methodology

## 2.1 Definition of Downward Trend

In this section, we will introduce the definition of a downward trend of the customer lifecycle. A customer typically shows a downward trend in several aspects. In an e-commerce company, gross merchandise volume (GMV), bought item count (BI) and purchase days (PD) all are meaningful metrics to track customer status. For each of these metrics, we define the downward flag respectively by the norm box method.

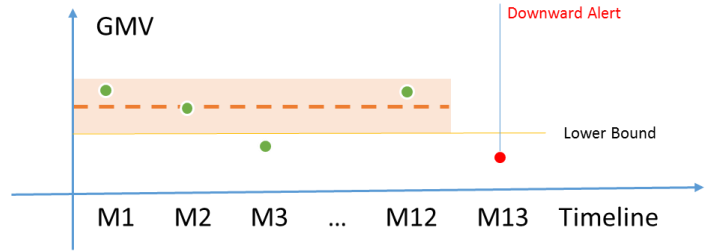

Fig 1. Downward Definition

Take the GMV as an example. Since consumption varies among customers, it is more appropriate to observe the GMV trend per user, rather than observing the GMV trend of the all customers. Hence, for each user, we first compute the average of the GMV $\mu$ in the past 12 months along with its standard deviation $s$ before a specified target time when we plan to communicate with the customer. In the next month, if the GMV of the customer is under the lower bound,

$$\mu - \alpha s$$

then we flag the user as being in a downward trend. $\alpha$ determines the sensitivity of the definition, i.e. a large $\alpha$ indicates that we are less likely to classify a user as in the downward trend; $\alpha$ should be tuned to match the business case and it is reasonable to use different $\alpha$ values for different customer groups. Figure 1 illustrates the method used to define the downward flag. The dashed line is the average GMV of a customer in the past 12 months. The orange box is the norm box corresponding to the upper bound $\mu + \alpha s$ and the lower bound $\mu - \alpha s$. If the user's GMV in the 13$^{th}$ month is lower than the lower bound, then the user is in a downward trend.

## 2.2 Modelling the Downward Propensity

Gradient Boosting Machine (GBM) is a widely used machine learning method for both regression and classification problems (Friedman, 2001). It shows strong performance in solving a wide range of practical applications (Natekin and Knoll, 2013). In this work, we leverage GBM to build binary classifiers to predict the downward trend of customers in the next month. If the user is in a downward trend per our definition, the class label is 1 (positive case); otherwise it is 0 (negative case). As mentioned in the previous section, three models are built for different downward detection goals, including GMV, BI and PD. The methodology is similar for the three metrics. Based on the purchase pattern of the customers, we divide the customers into frequent buyers (FB) and infrequent buyers (IB). We choose different norm box parameters $\alpha$ to define the downward for FB and IB. Event rate is the percentage of the positive case and obviously is decided by the $\alpha$. Generally, a proper $\alpha$ should be choose to make the event rate fit with the specific business case.

Table 1. Model Parameters

| Model | Event Rate | $\alpha$ (FB) | $\alpha$ (IB) |
|---|---|---|---|
| GMV | 9.65% | 1 | 0.75 |
| BI | 5.45% | 1.5 | 1 |
| PD | 7.32% | 1.25 | 1 |

We have around 200 candidate predictors belonging to multiple categories. We list some of them here:
- Transaction: purchase date, item price, category seasonality, condition, quantity, shipping methods.
- Bad buying experience history.
- Contact frequency and engagement
- Site behaviors: browse, bid, offer, watch, messages, cart, search, and dwell time.
- Demographics, acquisition channel

We first train rough GBM models using 200 trees and use a max depth of 5 for the 3 aspects. Based on the variable importance, we finalize the GBM model with 13 variables for each model in order to reduce the computing resource of the features while retaining acceptable prediction power. Table 2-4 shows the variables selected in each model.

Table 2. Variable Importance of GMV

| Variable Description | Importance |
|---|---|
| Whether last month's GMV is less than 0.75 std of previous 11 months' GMV | 1 |
| Recent 30 days' GMV ratio to past year | 0.4221 |
| Purchase days during past year | 0.3543 |
| Recent 180 days' GMV ratio to past year | 0.1746 |
| Average of gap between purchase days in last month | 0.0972 |
| Last 30 days' buying transactions ratio to past year | 0.0961 |
| Last 90 days' GMV ratio to past year | 0.0937 |

| Variable Description | Importance |
| --- | --- |
| Medium of gap between purchase days | 0.0928 |
| Last 60 days' GMV ratio to past year | 0.0845 |
| Standard deviation of gap between purchase days | 0.0549 |
| Whether last 3 months' GMV is less than 0.75 std of previous 9 months GMV | 0.0517 |
| User country | 0.0506 |
| Whether last 2 months' GMV is less than 0.75 std of previous 10 months GMV | 0.0361 |

Table 3. Variable Importance of BI

| Variable Description | Importance |
| --- | --- |
| Whether last month's bought items is less than 1 std of previous 11 months' bought items | 1 |
| Average of gap between purchase days in last month | 0.4387 |
| Medium of gap between purchase days | 0.3253 |
| Standard deviation of gap between purchase days | 0.2705 |
| Last 180 days' items purchased ratio to past year | 0.2588 |
| Purchase days during past year | 0.2407 |
| Last 30 days' items purchased ratio to past year | 0.2164 |
| Last 90 days' items purchased ratio to past year | 0.204 |
| Last year's items purchased count | 0.1049 |
| Whether last 3 months' bought items is less than 1 std of previous 9 months' bought items | 0.0776 |
| Last 30 days' buying transactions ratio to past year | 0.0757 |
| Last 60 days' items purchased ratio to past year | 0.0708 |
| User country | 0.0431 |

Table 4. Variable Importance of PD

| Variable Description | Importance |
| --- | --- |
| Purchase days during past year | 1 |
| Medium of gap between purchase days | 0.64 |
| Whether last month's purchase days is less than 1 std of previous 11 months' purchase days | 0.6221 |
| Average of gap between purchase days in last month | 0.3376 |
| Standard deviation of gap between purchase days | 0.1687 |
| Last 30 days' buying transactions ratio to past year | 0.1175 |
| Medium of purchase days difference in last month | 0.1003 |
| User country | 0.1001 |
| Standard deviation of purchase days difference | 0.0921 |
| Whether last month's bought items is less than 1 std of previous 11 months' bought items | 0.0908 |
| Last 180 days' buying transactions ratio to past year | 0.0837 |
| Average of purchase days difference | 0.0572 |
| Last 90 days' buying transactions ratio to past year | 0.0386 |

With the selected variables, we build GBM model for the three aspects using depths of 5

and 10. The detailed performance of the models can be found in Table 5. The performance is decent: the AUC is beyond 0.90 for all of these models. Figure 3 shows the log loss of GMV model along with the number of trees. 150 trees are efficient enough for prediction as the log loss no longer decreases dramatically. Comparing the 5-depth and 10-depth model, an obvious over-fitting issue can be observed in Figure 3 as the gap between the training and validation group. Meanwhile, there is no significant improvement with the 10-depth model. Thus, depth 5 model is selected as the final model. It is similar for the BI and PD models. Figure 3 shows the ROC curve and Precision-Recall (PR) Curve of the three final models.

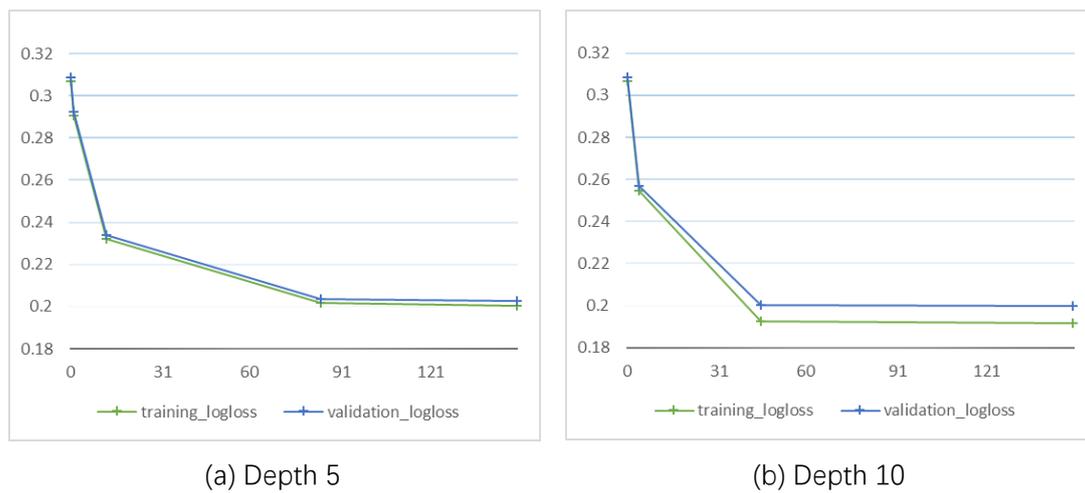

(a) Depth 5  (b) Depth 10

Figure 2. Log Loss of the GMV Model

In order to use the output of each model in practice, it is suggested to split the users into buckets according to the deciles or percentiles of the prediction score rather than set a fixed cut-off to assign positive or negative flags to the population. If bucket information is available, then it is feasible to combine these three models into one ensemble model to predict the downward trend. Each customer gets three buckets of the models output respectively, say $y_{GMV}, y_{BI}, y_{PD}$. If the goal is to find as many downward customers as possible, the $max(y_{GMV}, y_{BI}, y_{PD})$ can be selected as the final bucket of that customer.

Table 5. Model Performance Metrics

| Model | GMB | | BI | | PD | |
| --- | --- | --- | --- | --- | --- | --- |
| # Tree | 150 | 150 | 150 | 150 | 150 | 150 |
| depth | 5 | 10 | 5 | 10 | 5 | 10 |
| AUC | 0.90 | 0.90 | 0.91 | 0.91 | 0.90 | 0.91 |
| max f1 | 0.47 | 0.47 | 0.36 | 0.37 | 0.42 | 0.42 |
| max min per class accuracy | 0.82 | 0.82 | 0.83 | 0.83 | 0.82 | 0.82 |

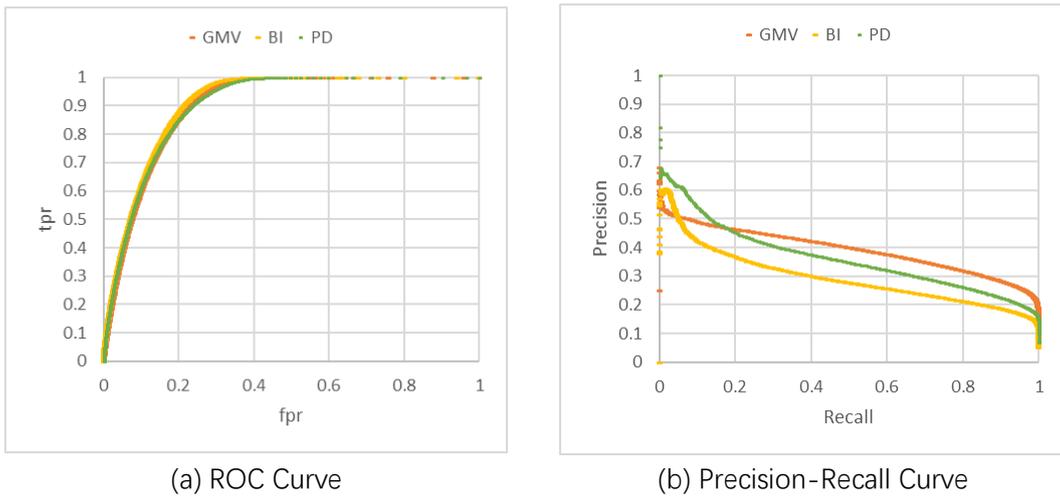

(a) ROC Curve  (b) Precision-Recall Curve

**Figure 2. Propensity Model ROC Curve and PR Curve**

## 2.3 Issue Detection

In this section, we will propose a methodology to identify one or more downward levers, which result in a downtrend trend. Generally, multiple causes lead a customer to a downward trend. To illustrate the method, we mainly focus on the BCE issue as an example, but it is convenient to apply the methodology to address other issues.

There are difficulties when considering some of the levers. Indeed, BCE issues are not negatively correlated with a downward trend in the whole population. The reason is that active customers tend to meet more BCE issues. The more transactions, the more chance a customer will have a BCE. Moreover, some of the customers are not bothered by the reported BCE because either the seller or customer service resolved the issue. Therefore, it is challenging but also crucial to find the real sufferers who behave downward due to a BCE. In addition, apart from customers who are willing to report their issues, there are a large group of BCE sufferers who choose not to speak out. Traditional business rules are not able to target them.

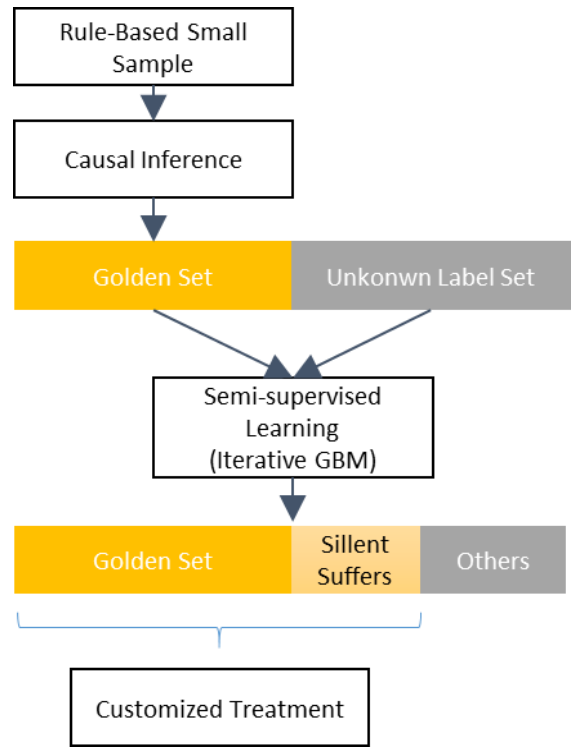

Fig 3. General Procedure of the Lever Detection Framework

In order to accomplish the goal, we need to resolve two questions. The first question is how to find the genuine sufferers from the BCE reporters. We name the genuine sufferer group as the golden set. Casual inference is applied to establish the golden set. The second question is how to detect the silent sufferers given the golden set; we use a semi-supervised learning method to deal with this. Figure 3 shows the general procedure of the proposed approach.

## 2.3.1 Casual Inference and Golden Set

To solve the first problem, we need to find a subset of the downward model population. The downward trend (GMV/BI/PD) of the customers in this subset is due to a BCE. Causal Inference is the theory of judging the relationship between two continuous variables. It helps to identify the cause or causes of a phenomenon (Shaughnessy and Zechmeister, 1985). For example, the method is used to detect casual signals within collections of static images (Lopez-Paz and Nishihara et al., 2016). In this work, we use the method to decide whether the BCE is the cause of the downward trend.

First, per business knowledge and sense, we start from a small customer sample who are in downward score decile 7-10 and had a BCE on their last purchase day; we assume that they are real sufferers.

Second, we try to demonstrate causality. Let X be the cumulative BCE count in the past; Y be the model decile (higher decile, higher likelihood to be downward). We do causal inference separately corresponding to the customer type, as the transactional behaviors of FB and IB are quite different. It is expected to observe asymmetry where X can cause Y but Y cannot cause X.

Table 6. Casual Inference

| Relationship | Coefficient | With CL $\alpha = 0.001$ |
|---|---|---|
| **Frequent Buyer** | | |
| $Y \leftarrow f(X) + e$ | 0.008 | Pass |
| $X \leftarrow g(Y) + e$ | 0.53 | Fail |
| **Infrequent Buyer** | | |
| $Y \leftarrow f(X) + e$ | 0.03 | Pass |
| $X \leftarrow g(Y) + e$ | 0.007 | Fail |

We list the linear regression results in the Table 6. It indicates that the asymmetry exists and the initial sample can be considered as Golden Set. For FB, with the confidence level 0.001, the casual inference suggests that the BCE is the cause of the downward trend; but the BCE reports is not the consequence of the downward trend. Similar conclusion can be induced for IB.

## 2.3.2 Semi-supervised Learning and Look-like Model

We conclude that the customers in the Golden Set are the ones who suffered from BCE by causal inference. However, there must be additional customers who suffered from BCE but they did not report the issue. We here name them as silent sufferers. In this section, semi-supervised learning theory is applied to find the silent sufferers. In some of the cases, labeled data is not enough to build a supervised learning model and handling the situation semi-supervised learning is a potential solution (Zhu, 2005). For our case, we have a limited golden set; the rest are unknown labeled customers. An expectation–maximization (EM) algorithm is used in this work (Liu and Lee et al., 2002).

The label of the Golden Set in a supervised learning fashion can be positive (i.e. 1). However, for the remainder of the users, we do not know their labels. Some of them are silent sufferers who should be also labeled as 1, while others are "normal" users and their labels should be 0. In order to label these unknown customers, semi-supervised learning technique is used in this section to solve the problem.

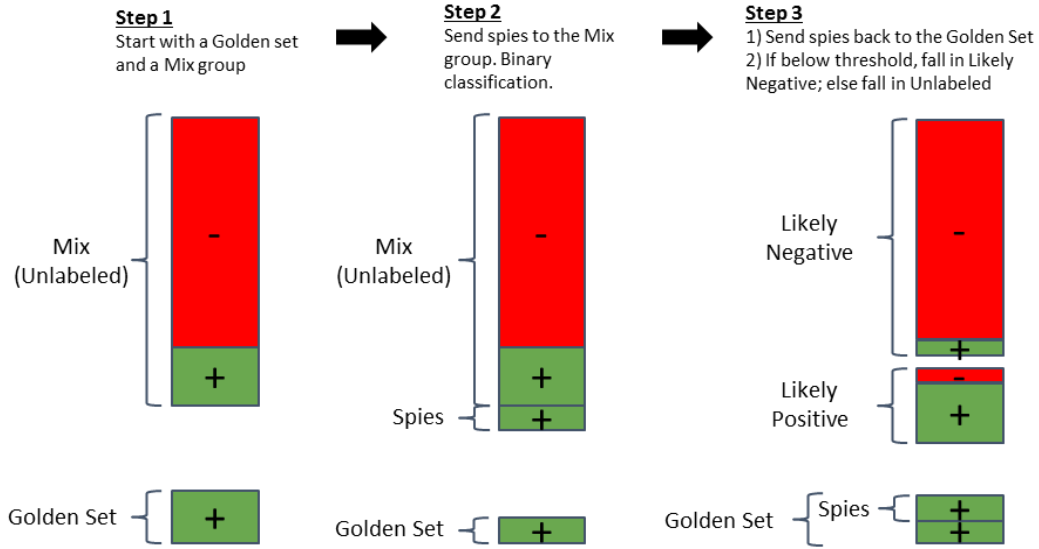

Fig 4. Initialize the Semi-supervised Learning

| Algorithm 1 |
|---|
| Initialize parameters. Set label $y_i$ of golden set $G = \{(x_i, y_i)\}$ as 1; mixed data set $M = \{(x_i, y_i)\}$ as 0. Set the stop criteria $\theta$. Set the change rate $\beta = 1$. Set the maximum iteration $N$. <br><br> While $\beta > \theta$ and $k < N$ <br> 1 Randomly picking a subset $G' \subset G$ set as spies. <br> 2 Build binary classifier $f_k(x)$ on the dataset $M' = G' \cup M$. <br> 3 Scoring the combined dataset $M'$ and update the labels on $M'$. <br> 4 Send the spies back to golden set and re-labeled them as 1. <br> 5 $M = M'\backslash G'$ <br> 6 Compute the label change rate $\beta$ on $M$. <br> 7 $k = k + 1$ |

Figure 4 illustrates the one iteration step of the learning procedure. In step 1, the Golden Set and the remainder of the unknown dataset are mixed without labels. We then set the initial labels of the unlabeled customers as all 0s. In step 2, we randomly select part of the positive samples as spies and combine with the mixed part. It is now feasible to train a supervised learning model to get a binary classifier, although it is biased due to the unknown labels of the mixed dataset. In step 3, send all the spies back to the Golden Set. For the remaining mixed part, use the binary classifiers to re-label all the samples utilizing a specified cut-off. Notice that after re-labeling, some of the samples change their labels. The overall label change rate can be referred as stop criteria. When the change rate is lower than the threshold, stop the iteration procedure. The detailed algorithm is organized in the

To build the binary classifier in step 3, the variables are from the following aspects: BCE reports, customer service (call, online chat, email), help & FAQ page behaviors, buyer's feedback to sellers, survey data, and behavioral features. The cut-off is selected by the max $F_1$ scores in Step 3. F-score is used to trade off the precision and recall. It is possible to choose other cut-off such as $F_{0.5}$, $F_2$ to fit with a different application (Li and Wang et al., 2008).

## 2.3.3 Evaluation

It is difficult to tell the model performance on the mixed dataset since the true labels are all unknown. Nonetheless, the performance on the Golden Set still can be observed as a reasonable reference. In an ideal case, the model should classify all customers in the golden set as positive cases. We divide the golden set into two parts: one part is combined with the mixed set as a training set; the other is a holdout validation set. We check the accuracy on the hold out validation set via the binary classifier built in each iteration. We consider the accuracy as recall since it is similar to the original meaning of the concept. It is expected that the recall improves along with iteration. In addition, the label change rate on the mixed dataset should be stable after a sufficient number of iterations.

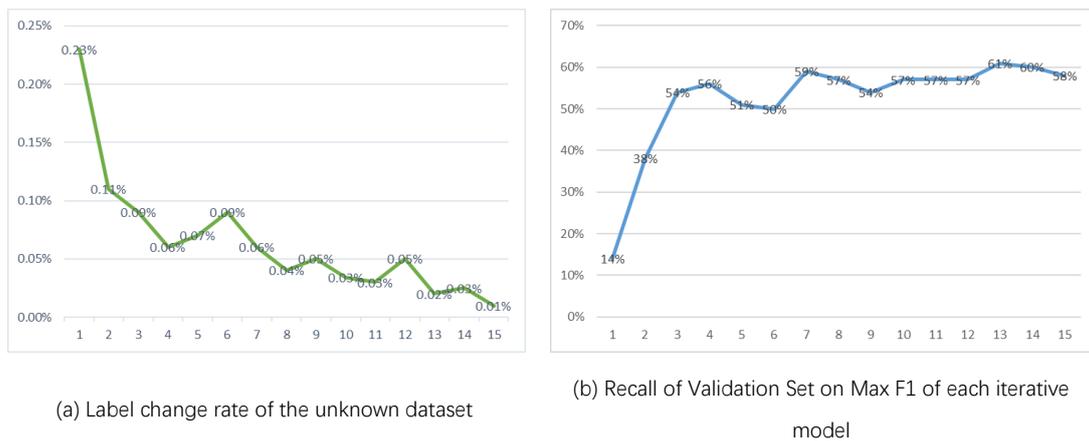

(a) Label change rate of the unknown dataset

(b) Recall of Validation Set on Max F1 of each iterative model

**Fig 6. Iteration Procedure of Trust Lever**

Figure 6 shows the iteration of the algorithm. As the recall tends to be stable and the label change rate is near to zero, we can conclude that iteration 11 is an ideal choice as the final model. In Table 7, we list the top five variables of the selected binary classifier.

Table 7. Top five variables in the Trust Lever Semi-supervised Learning Model

| Variable Name | Variable Type |
| --- | --- |
| # of days since last defect | BCE |
| Defect rate of pre 7 days(without late delivery) | BCE |
| Defect rate of pre 1 year(without late delivery) | BCE |
| BCE count of pre 1 year | BCE |

| # of days since last purchase | Transaction |
|---|---|

# 3. Model Performance

In this section, we set up test to verify the correctness of the population selected by our framework and influence of the model in real business case. As the model targets the downward users caused by BCE, it is reasonable to reach them with an apology message about the customer experience. We launched an A/B campaign test in US and UK in Sep. 2017. Using the labels from model prediction, we chose 1,012 customers who were in the Golden Set or Silent Sufferers set. Next, we randomly separated those into test and control groups. For the test group, we sent each one an email to apologize. Details of the population can be found in Table 8.

Table 8. Settings of the Campaign

|  | # Golden Set | # Silent Suffers | Grand Total |
|---|---|---|---|
| Control | 143 | 55 | 198 |
| Test | 572 | 242 | 814 |
| Grand Total | 715 | 297 | 1,012 |

After the campaign run date, the lift of GMV in the test group is 88.5% comparing with the control group. Through a two-sample t-test, which is used to determine if the means of two groups are equal (Jones, 1994), it suggests that the improvement of the test group comparing with the control group is significant with the confidence level $\alpha = 0.05$. The decent lift indicates that we targeted the right population and got their positive feedback.

# 4. Discussion

In this work, we propose a scientific framework to focus on the downward trend customers in the business, leveraging multiple machine-learning techniques. We first propose a method to define and detect the propensity of downward trend of customers, which becomes the foundation for the following step. Next, casual inference and semi-supervised learning are applied to find golden set and silent sufferers. In an A/B campaign test, we verify the performance of the model and ensure the effectiveness of the model.

The casual inference and semi-supervised learning part are adapted to other stages of customer lifecycle as well, including churn, early churn and one-and-only-one-transaction cases. For lever detection in this study, we build and test the frame on the BCE lever. However, other levers such as spend capacity are also worth efforts. Moreover, we introduce the methodology in an e-commerce business background. It is effortless to apply in other business areas. For instance, in the telecommunication industry, we can use the variables such as call quality, device type, and count of the text message sent among others to develop the framework. We will focus on improving performance and extend to other CRM application scenarios.